\documentstyle[12pt,A4]{article}
\begin{document}
\label{zacatek-vys4}
\noindent
\hspace*{8cm}{\sc acta univ. palacki. olomuc.,}\\
\hspace*{8cm}{\sc fac.\,rer.\,nat.\,{\small (1998)},\,physica\,{\small
37}},\\
\hspace*{8cm}{\small \pageref{zacatek-vys4} -- \pageref{konec-vys4}}\\
\hspace*{8cm}\rule[3mm]{6.2cm}{0.2mm}\\[1cm]
\centerline{\large \bf NOTE ON THE CHANDRASEKHAR MODEL}\\
\centerline{\large \bf OF THE OPTICAL ACTIVITY OF CRYSTALS}\\[2mm]
\centerline{\bf Jan \v R\'\i ha, Kamila Sv\' a\v ckov\' a, Ivo Vy\v s\'\i n} \\[3mm]
\noindent {\small Department of Theoretical Physics, \ Natural Science
Faculty of Palack\' y University, \ Svobody 26, {\bf 771~46 Olomouc}, \
Czech Republic}\\[4mm]
\centerline{\it Received 26th November 1997}

\vspace*{0.5cm}
%----------------------------------------------------------------------------
%-----------------------------------------------------------------------------
{\sc ABSTRACT:} \ In this paper we discuss more wide applicability
of the Chandrasekhar model of coupled oscillators in the optical
rotatory dispersion of the crystals. We solve the problem using
the Chandrasekhar model of two coupled oscillators in the case when we
include all couplings between adjacent oscillators on the helix that is
given by the crystal structure. Further we discuss the results of
coupled oscillators models in the case of the including of the couplings
between even and odd oscillators on the helix. The ORD results obtained
after the approximations of the oscillator strengths verify that these
couplings have the important influence on the crystal ORD.
%-------------------------------------------------------------------------------
\section{Introduction}
\label{uvod}
\hspace \parindent
%------------------------------------------------------------------------------
The phenomenon of the optical activity (OA) has two aspects which arise
from  the interaction of the radiation with matter - dispersive and absorptive.
These aspects are the optical rotatory dispersion (ORD) and the circular
dichroism (CD). It is well known that ORD is the dependency of the
rotation of the linear polarized light per unit length on the frequency
$\omega $ or on the wave length $\lambda $. The CD is the ellipticity
per unit length of the wave getting off the crystals. In the difference from the ORD,
which is nonzeroth in a wide frequency region, the CD is nonzeroth
only in a very narrow frequency region in the absorption region.

The important group of optically active crystals are the crystals with
screw axis of symmetry belonging to the space groups of symmetry
$D_{3}^{4}$ and $D_{3}^{6}$. The typical representatives of these
crystals are $\alpha $-quartz, cinnabar, tellurium, selen, camphor,
benzil. The optical activity of these crystals due to asymmetrical
originating of the crystal structure because the molecules or atoms
forming the crystals are not optically active. Camphor is the exception
because his molecules are optically active so that his optical activity has
two sides - molecular and crystalline.

In the past the OA of crystals was studied by more authors and their works are
based on the different theories. We can introduce the theory of excitons
\cite{Agr1,Agr2,Tsv,Kato}, the theory of coupled oscillators
\cite{Chand1,Chand2,Vys1} or Lagrangian formalism \cite{Nels}. But it is
known that the theory of coupled oscillators gives the results which are
very good applicable in the fitting of the experimental data
of OA. The model of coupled oscillators was for the first time used by
Chandrasekhar \cite{Chand1} which has applicated the Kuhn model
\cite{Kuhn} of two coupled oscillators which represent the smallest unit
of the optical active crystal. He has used the following model: The first
oscillator
lies in the plane $z=0$ and its position is given by direction cosines
$\alpha ,\beta ,\gamma $. The second oscillator lies in the plane $z=d$
and its direction of vibration is turned by the angle $\theta $ around
the $z$ axis which is parallel to the crystal axis $c$. Both oscillators lie
on the helix which is given by the crystal structure. The $z$ axis
(crystal axis $o$) has the direction of the propagation of the linear
polarized electromagnetic wave. Both coupled oscillators forming one
compound oscillator are identical. We denote the coupling constant between
oscillators as $Q_{1}$. Chandrasekhar further has assumed that
the number of compound oscillators in the volume unit is $N/2$, where
$N$ is the number of single oscillators in the volume unit. The
influence of the interactions between oscillators belonging to the other
helices is neglected because of another type of couplings.

Becouse of the interaction between single oscillators in the compound
oscillator the natural frequency $\omega _{0}$ is split into two frequencies
$\omega _{1}$ and $\omega _{2}$ of normal modes of vibrations. For these
normal modes Chandrasekhar has solved the dispersion theory of
refractive indices for the propagation of the left and
the right circularly polarized wave into which the linear polarized wave is
split in the optically active medium. It is well known that the medium
containing coupled oscillators is optically active and therefore it must be
characterized by the different refractive indices $n_{l}$ and $n_{r}$
for the left and the right circularly polarized wave. Chandrasekhar has solved
the dispersion theory semiclasically, he has introduced into results the
oscillator strengths but he has assumed that the oscillator strengths of
the normal modes of vibrations are the same.

Chandrasekhar has solved his model of optical activity for the $\alpha
$-quartz. Because the CD is not measured for the $\alpha $-quartz and
therefore nor the ORD in the absorption region Chandrasekhar has
assumed the oscillators as undamped.

Later V. Vy\v s\'\i n
\cite{Vys1} has removed Chandrasekhar's simplifications in the approach
of the oscillator strengths of the normal modes of vibrations and he has
obtained for the ORD which is denoted by $\rho (\omega )$ the formula
%------------------------------------------------------------------------------
\begin{equation}
\label{1}
\rho (\omega )=\frac{\pi Nde^{2}\omega ^{2}}{mc^{2}}\left( \alpha ^2+\beta
^2\right) \sin \theta \left[ \frac{-f_{g_{1}}}{\omega _{1}^{2}-\omega
^{2}}+\frac{f_{q_{2}}}{\omega _{2}^{2}-\omega ^{2}}\right],
\end{equation}
%------------------------------------------------------------------------------
where $f_{q_{1}}$ and $f_{q_{2}}$ are the oscillator strengths of the normal
modes of vibrations, $m$ is the mass of the single oscillator, $e$ his
electric charge and $c$ is the velocity of the light. The frequencies of
the normal modes are $\omega _{1}^{2}=\omega _{0}^{2}+Q_{1}$, $\omega
_{2}^{2}=\omega _{0}^{2}-Q_{1}$. Because the splitting of the frequencies of
the normal modes is very small we can (\ref{1}) rewrite in the form
containing the natural frequency of single oscillators
%------------------------------------------------------------------------------
\begin{equation}
\label{2}
\rho (\omega )=\frac{\pi Nde^{2}\omega ^{2}}{mc^{2}}\left( \alpha ^2+\beta
^2\right) \sin \theta \cdot \left[ \frac{f_{q_{2}}-f_{g_{1}}}{\omega _{0}^{2}-\omega
^{2}}+\frac{Q_{1}\left( f_{q_{1}}+f_{q_{2}}\right) }{\left( \omega
_{0}^{2}-\omega ^{2}\right) ^{2}}\right].
\end{equation}
%-------------------------------------------------------------------------------

We see that the ORD based on the two coupled oscillator model leads
in general to the two member formula. The first term on the right side
of eq. (\ref{2}) is known as Drude's term and the second term is known as
Chandrasekhar's term. This is problem because we don't know the oscillator
strengths for the real crystals. We can solve the oscillator strengths
only in approximations. For example in the approximation of the
linear harmonic oscillator $f_{q_{1}}=f_{q_{2}}=f_{0}$ (the Chandrasekhar
approximation) the formula (\ref{2}) leads only to the Chandrasekhar formula
%-------------------------------------------------------------------------------
\begin{equation}
\label{3}
\rho (\omega )=\frac{2\pi Nde^{2}Q_{1}f_{0}}{mc^{2}}\left( \alpha ^2+\beta
^2\right) \sin \theta \cdot \frac{\omega ^{2}}{\left( \omega
_{0}^{2}-\omega ^{2}\right) ^{2}}.
\end{equation}
%--------------------------------------------------------------------------------

On the other hand we can solve the oscillator strengths in the Heitler -
London approximation \cite{Vvjc} which gives the result
%-------------------------------------------------------------------------------
\begin{equation}
\label{4}
\frac{f_{q_{1}}}{\omega _{1}}=\frac{f_{q_{2}}}{\omega
_{2}}=\frac{f_{0}}{\omega _{0}}
\end{equation}
%-------------------------------------------------------------------------------
and then ORD is given by the formula
%-------------------------------------------------------------------------------
\begin{equation}
\label{5}
\rho (\omega )=\frac{\pi Ne^{2}dQ_{1}f_{0}}{mc^{2}}\left(
\alpha ^2+\beta
^2\right) \sin \theta \cdot \frac{\omega ^{2}\left( \omega _{0}^{2}+\omega
^{2}\right)}{\omega _{0}^{2}\left( \omega
_{0}^{2}-\omega ^{2}\right) ^{2}}.
\end{equation}
%--------------------------------------------------------------------------------

We note that the formula (\ref{5}) is the Agranovich type
\cite{Agr1,Agr2} which was derived by the exciton's theory.

The formula (\ref{2}) or its special forms (\ref{3}) and
(\ref{5}) was used in many works to describe the experimental data
of ORD of the crystals, for example of the tellur
\cite{Nom,Ades,Fukuda}, of the $\alpha $-quartz \cite{Buhr,Burkov} or of
the crystal $Bi_{12}GeO_{20}$ \cite{Hen1, Hen2} and without exception
with good results. On the other hand we think that the Chandrasekhar
model still contains some simplifications which we want to solve in this
paper.

The Chandrasekhar model assumes that $N/2$ coupled oscillators, where $N$ is
the number of single oscillators, exist in the volume unit. But
it means that we have $N/2$ isolated compound oscillators in the volume unit.
This idea desribes for example the system of randomly
oriented isolated non-interacting molecules. But in the crystals
all adjacent molecules or atoms interact on the helices. It means that
the second oscillator in the any compound oscillator is at the same time
the first oscillator in the next compound oscillator. It is the first
question how the results of the Chandrasekhar model will change in the case
when we include all couplings between all adjacent oscillators into this
model.

The second question follows from the fact that the two coupled
oscillators model neglect all other couplings between oscillators on
one helix. It is possible in the case when all oscillators lie on
the line because the value of the coupling constant decreases with third
power of the distance. But it is not true in the real crystals where
the diameter and the step of the helix are comparable. In these cases we
must include also the couplings between even and odd oscillators on the
helices. Further in the crystals with the space groups of symmetry
$D_{3}^{4}$ and $D_{3}^{6}$ the vibration directions of the adjacent
oscillators contain the angle $\theta =120deg$. The vibration directions of
the even or odd oscillators contain the angle $\theta
=240deg$. From the formulae (\ref{2}), (\ref{3}) or (\ref{5}) we see
that the sense of ORD depends on the value of $\sin \theta$.
It means that the
couplings between even and odd oscillators on the helices have the
opposite effect on the ORD in compare with the couplings between adjacent
oscillators.
We note that in the crystals with space groups of symmetry
$D_{3}^{4}$ and $D_{3}^{6}$ the couplings between the first and the
fourth oscillators which are randomly oriented don't have the influence
on the OA and then the other couplings are really neglectable.

%-------------------------------------------------------------------------------
\section{The influence of the couplings between all adjacent
oscillators}
\label{kap2}
\hspace \parindent
We can discuss this problem by means of the results of our previous
paper \cite{Vys3}. In this paper we have solved the application of the
ORD Chandrasekhar model of three coupled oscillators. We have added to
the Chandrasekhar model of two coupled oscillators the third
oscillator which lies on the helix in the plane $z=-d$ and its
vibration direction is turned by the angle $-\theta $ around $z$ axis with
respect to the oscillator in the plane $z=0$. We include also as
Chandrasekhar only the couplings between adjacent oscillators. The
number of these compound oscillators in the volume unit is $N/3$.

The solving method is also the same as Chandrasekhar's with the
exception that the natural frequency $\omega _{0}$ of each oscillators
is split into three frequencies $\omega _{1}$, $\omega _{2}$ and $\omega
_{3}$ and also the ORD result contains three terms regards to (\ref{1})
that is
%--------------------------------------------------------------------------
\begin{eqnarray}
\label{6}
\rho (\omega )&=&\frac{2\pi Nde^{2}\omega ^{2}}{3mc^{2}}\left( \alpha
^{2}+\beta ^{2}\right) \sin \theta \nonumber \\
&&\times \left[ \frac{-\left( \sqrt{2}+2\cos \theta \right)
f_{q_{1}}}{\omega _{1}^{2}-\omega ^{2}}+\frac{4f_{q_{2}}\cos \theta
}{\omega _{2}^{2}-\omega ^{2}}+\frac{\left( \sqrt{2}-2\cos \theta \right)
f_{q_{3}}}{\omega _{3}^{2}-\omega ^{2}}\right] ,
\end{eqnarray}
%---------------------------------------------------------------------------
where $\omega _{1}^{2}=\omega _{0}^{2}+\sqrt{2}Q_{1}$, $\omega
_{2}^{2}=\omega _{0}^{2}$ and $\omega _{3}^{2}=\omega
_{0}^{2}-\sqrt{2}Q_{1}$. We can rewrite this result with the natural
frequency of oscillators $\omega _{0}$ but this result is involved too.
We express this result only in the approximations of the oscillators
strengths. In the Chandrasekhar linear harmonic oscillator approximation
$f_{q_{1}}=f_{q_{2}}=f_{q_{3}}=f_{0}$ we get
%-----------------------------------------------------------------------------
\begin{equation}
\label{7}
\rho (\omega )=\frac{8\pi Nde^{2}Q_{1}f_{0}\omega ^{2}}{3mc^{2}}\left( \alpha
^{2}+\beta ^{2}\right) \sin \theta \cdot \left[ \frac{1}{\left( \omega
_{0}^{2}-\omega ^{2}\right) ^{2}}-\frac{2Q_{1}\cos \theta }{\left(
\omega _{0}^{2}-\omega ^{2}\right) ^{3}}\right].
\end{equation}
%-------------------------------------------------------------------------------

We see that the second term in the square brackets is very small with
respect to the first one. It contains the coupling constant $Q_{1}$ that is
very small and the denominator of this term is in the frequency region
far from absorption much greater than the denominator of the first term.
From this reason the second term can be neglected.

In the Heitler - London approximation that is given by the extended
relation (\ref{4}) we have
%------------------------------------------------------------------------------
\begin{equation}
\label{8}
\rho (\omega )=\frac{4\pi Nde^{2}Q_{1}f_{0}\left( \alpha ^{2}+\beta
^{2}\right) \sin \theta }{3mc^{2}}\cdot \left[ \frac{\omega ^{2}\left(
\omega _{0}^{2}+\omega ^{2}\right) }{\omega _{0}^{2}\left( \omega
_{0}^{2}-\omega ^{2}\right )^{2}}-\frac{2Q_{1}\cos \theta }{\left( \omega
_{0}^{2}-\omega ^{2}\right) ^{3}}\right]
\end{equation}
%-------------------------------------------------------------------------------
and we can neglect again the second term in the square brackets. But
then the results of two and three coupled oscillators are very similar,
they differ only in the multiplicative constant $4/3$. We see the sense of
this constant if we compare the results for one compound oscillator.
We must divide the results of two coupled oscillators model (\ref{3}) and
(\ref{5}) by $N/2$
(the number of compound oscillators in the volume unit) and the results
of the three coupled oscillators model (\ref{7}) and (\ref{8}) by $N/3$.
By the comparing of corresponding results we see that
one triad of the coupled oscillators has the twofold effect on the ORD
than one couple. From this it follows that the effects of couplings
between adjacent oscillators are aditive. If we want to include all
couplings in the two coupled oscillators model we must take as the number
of coupled oscillators $N$ (in the case $N\rightarrow \infty$)
because the number of couplings is the same as the number of single
oscillators. In the case of three coupled oscillators model we take N/2
as the number of compound oscillators in the volume unit etc. Then both
models of two and three coupled oscillators give the same results in
the case of the including of the couplings between adjacent oscillators
only. The mistake of this conclusion is given by the neglecting of the
second terms in the square brackets in (\ref{7}) and (\ref{8}) but
we can assume that in the cases of small couplings in the crystals the
mistake is neglectable.

From our conclusions in this section it follows that if we use our
results in the approximation of experimental data of ORD of the
crystals we get different values of unknown parameters in these
formulae. For example the ORD of $\alpha $-quartz was approximated by
Chandrasekhar's and Vy\v s\'\i n's results of the two
coupled oscillators model. The unknown parameters were the oscillator
strengths of the normal modes of vibrations. The values of the oscillator
strengths were in the results that are assumed as true smaller as the
number of valence shell electrons \cite{Vys4}. In the other results the
values of the oscillator strengths were greater. But after our correction
of the results of the two coupled oscillators model we get for all
oscillator strengths a half of values and it would be difficult to say what
results of the approximations of the experimental data are true and what the
values of all oscillator strengths will be too small. It is the next
reason for the assertion that we must include also the couplings between
even and odd oscillators which due to its opposite effect on the ORD
return for example in the case of $\alpha $-quartz the results of the
approximations of oscillator strengths to the values that they have in
\cite{Vys4}.
%---------------------------------------------------------------------------
\section{The influence of the couplings between even and odd oscillators}
\label{sec3}
\hspace \parindent
%---------------------------------------------------------------------------
The smallest system in which we can include the coupling between even or
odd oscillators in the helix is the system of three coupled oscillators
if we include also the coupling between the first and the third
oscillator. The scheme of this system is the same as in the previous
section. We characterize the coupling between the first and the third
oscillator by the coupling constant $Q_{2}$. Regarding to the previous
section we use now the value $N/2$ as the number of compound
oscillators where $N$ is the number of single oscillators in the volume
unit. It means that the last single oscillator in any compound
oscillator is the first single oscillator in the next compound
oscillator on the helix.

The solving ORD of this system by the Chandrasekhar semiclassical
model is described in \cite{Vys5}. We get for the ORD the three member
formula
%----------------------------------------------------------------------------
\begin{eqnarray}
\label{9}
\rho (\omega )&=&\frac{4\pi Nde^{2}\omega ^{2}}{mc^{2}}\left( \alpha
^{2}+\beta ^{2}\right)
\sin \theta \nonumber \\
&&\times \left[ \frac{\frac{-\left(2\cos \theta +A_{1}\right)
}{2+A_{1}^{2}} f_{q_{1}}}{\omega _{1}^{2}-\omega
^{2}}+\frac{f_{q_{2}}\cos \theta }{\omega _{2}^{2}-\omega ^{2}}+
 \frac{\frac{-\left(2\cos \theta +A_{3}\right)
}{2+A_{3}^{2}} f_{q_{3}}}{\omega _{3}^{2}-\omega
^{2}}\right] ,
\end{eqnarray}
%-------------------------------------------------------------------------------
where $\omega _{1}^{2}=\omega
_{0}^{2}+\frac{Q_{2}+\sqrt{8Q_{1}^{2}+Q_{2}^{2}}}{2} $, $\omega
_{2}^{2}=\omega _{0}^{2}-Q_{2}$ and $\omega _{3}^{2}=\omega
_{0}^{2}+\frac{Q_{2}-\sqrt{8Q_{1}^{2}+Q_{2}^{2}}}{2} $ and $A_{1}$,
$A_{3}$ are $A_{1}=\frac{-Q_{2}+\sqrt{8Q_{1}^{2}+Q_{2}^{2}}}{2Q_{1}}$,
$A_{3}=\frac{-Q_{2}-\sqrt{8Q_{1}^{2}+Q_{2}^{2}}}{2Q_{1}}$. We see
that by the expression of this result regarding to the natural frequency
we obtain complicated result and its form is also described in
\cite{Vys5}. Let us again be interested in the form of the result after
the approximations of the oscillator strengths. If we neglect as in previous
section all the terms
containing $1/(\omega _{0}^{2}-\omega ^{2})^{3}$ we get in the Chandrasekhar
approximation the result
%------------------------------------------------------------------------------
\begin{equation}
\label{10}
\rho (\omega )=\frac{4\pi Nde^{2}f_{0}
\left( \alpha
^{2}+\beta ^{2}\right) \sin \theta \cdot \left( Q_{1}+2Q_{2}\cos
\theta \right)
}{mc^{2}} \cdot \frac{\omega ^{2} }{\left( \omega _{0}^{2}-\omega ^{2}\right)
^{2}}
\end{equation}
%------------------------------------------------------------------------------
and in the Heitler - London approximation
%-------------------------------------------------------------------------------
\begin{equation}
\label{11}
\rho (\omega )=\frac{2\pi Nde^{2}f_{0}
\left( \alpha
^{2}+\beta ^{2}\right) \sin \theta \cdot \left( Q_{1}+2Q_{2}\cos
\theta \right)
}{mc^{2}} \cdot \frac{\omega ^{2}\left( \omega _{0}^{2}+\omega
^{2}\right) }{\omega _{0}^{2}\left( \omega _{0}^{2}-\omega ^{2}\right)
^{2}}.
\end{equation}
%------------------------------------------------------------------------------

If we compare these results with the results (\ref{7}) and (\ref{8}) after
their discussion, it means after neglecting the terms with $1/(\omega
_{0}^{2}-\omega ^{2})^{3}$ and after their correction by including all
couplings between adjacent oscillators (the coefficient $N/3$ is
substituted by $N/2$), we see that the results (\ref{10}) and
(\ref{11}) are different only in
the coefficient $Q_{1}+2Q_{2}\cos \theta $ which substitutes the coupling
constant $Q_{1}$ in the eqs. (\ref{7}) and (\ref{8}). In the crystals
belonging to the space groups of symmetry $D_{3}^{4}$ and $D_{3}^{6}$ is
$\cos \theta =-0.5$ and the coefficient gives the value $Q_{1}-Q_{2}$.
The couplings between even and odd oscillators have indeed the opposite
efect on the ORD. Besides we include in our three coupled oscillators
model all couplings between adjacent oscillators on the helices
but only a half of couplings between even and odd oscillators. Nevertheless
the influence of couplings with couplings constants $Q_{1}$ and $Q_{2}$
has the same form. If we include all couplings between even and odd
oscillators the influence of these couplings will be, of course, much
greater but we don't know the expression for this influence.

Even and odd coupled oscillators on the helices are not contacted
and we cannot use the method of the contacted compound oscillators
from the section \ref{kap2}. If we want to solve this problem exactly we
should solve the model of $\bar N$ coupled oscillators where $\bar N$ is
the number of all single oscillators on one helix and in this model we
should
include all couplings between adjacent, even and odd oscillators. This
model is not, of course, really solvable.

From this reason we have proceeded by another way. We have solved
further the models of four, five etc. coupled oscillators in which we
have included the couplings
between adjacent, even and odd oscillators and the results we have
expressed only with the natural frequency $\omega _{0}$ and only in the
approximations of the oscillator strengths of the normal modes of vibrations
and only for the crystals with the space groups of symmetry $D_{3}^{4}$ and
$D_{3}^{6}$.
Based on these results we have found the tendency in the
expression that contains the coupling coefficient $Q_{2}$ and we have
looked the limit of this expression with increased number of oscillators
in the models. We show now the solving of the four coupled
oscillators model. The solving of other models is similar but more
complicated and we discuss only the results.

%--------------------------------------------------------------------------
\subsection{The four coupled oscillator model of the optical rotatory
dispersion of crystals}
\label{sec4}
\hspace \parindent
%--------------------------------------------------------------------------

We extend the three coupled oscillators model by the fourth
oscillator which lies on the helix in the plane $z=2d$ and his vibration
direction is turned by the angle $2\theta $ around $z$ axis with
respect to the vibration direction of the oscillator in the plane $z=0$.
The coupling constant $Q_{1}$ describes the coupling between adjacent
oscillators and the constant $Q_{2}$ describes the couplings between
the first and the third and between the second and the fourth
oscillators. The motion equations of the oscillators in the fields of
the left and the right circularly polarized wave are
%---------------------------------------------------------------------
\begin{equation} \label{pohybove rovnice}
\begin{array}{c}
\ddot r_{1} + \omega_{0}^{2}r_{1} + Q_{1}r_{2} + Q_{2}r_{3} =
F_{1}^{l,r}, \\
\ddot r_{2} + \omega_{0}^{2}r_{2} + Q_{1}r_{1} + Q_{1}r_{3} + Q_{2}r_{4}
= F_{2}^{l,r}, \\
\ddot r_{3} + \omega_{0}^{2}r_{3} + Q_{2}r_{1} + Q_{1}r_{2} + Q_{1}r_{4}
= F_{3}^{l,r}, \\
\ddot r_{4} + \omega_{0}^{2}r_{4} + Q_{2}r_{2} + Q_{1}r_{3} =
F_{4}^{l,r},
\end{array}
\end{equation}
%-------------------------------------------------------------------------
where $r_{1}$, $r_{2}$, $r_{3}$, $r_{4}$ are the displacements of
oscillators from equilibrium, $F$ denotes the projection of functioning
forces to the motion direction of oscillators divided by the masses
$m$ of oscillators (electrons). In all terms in equation (\ref{pohybove rovnice}) and further
in the sign $\pm$ hold the $+$ sign for the left and the $-$ sign for the
right cicularly
polarized wave. The electric field vector $\vec E$ has in our case the
components
%--------------------------------------------------------------------------
\begin{equation} \label{slozky elektricke intenzity}
\begin{array}{c}
E_{x}^{l,r} = E_{0}\cos(\omega t - k_{l,r}z),\\
E_{y}^{l,r} = \pm E_{0}\sin(\omega t - k_{l,r}z)
\end{array}
\end{equation}
%-------------------------------------------------------------------------
and then we can express the forces $F_{\eta}^{l,r} (\eta=1,2,3,4)$ as
%-------------------------------------------------------------------------
\begin{equation} \label{sily oscilatoru 1}
\begin{array}{c}
F_{1}^{l,r} = \frac{eE_{0}}{m}[(\alpha\cos\theta + \beta\sin\theta) \cos(\omega t +
\phi_{l,r}) \\
\pm (- \alpha\sin\theta + \beta\cos\theta) \sin(\omega t + \phi_{l,r})],\\
F_{2}^{l,r} = \frac{eE_{0}}{m}(\alpha\cos\omega t \pm \beta\sin\omega t),\\
F_{3}^{l,r} = \frac{eE_{0}}{m}[(\alpha\cos\theta - \beta\sin\theta) \cos(\omega t -
\phi_{l,r}) \\
\pm (\alpha\sin\theta + \beta\cos\theta) \sin(\omega t - \phi_{l,r})],\\
F_{4}^{l,r} = \frac{eE_{0}}{m}[(\alpha\cos 2\theta - \beta\sin 2\theta) \cos(\omega t -
2\phi_{l,r}) \\
\pm (\alpha\sin 2\theta + \beta\cos 2\theta) \sin(\omega t -
2\phi_{l,r})],
\end{array}
\end{equation}
%---------------------------------------------------------------------------
where $e$ is the electron charge and $\phi_{l,r} = k_{l,r} = \frac{2\pi
n_{l,r}d}{\lambda} = \frac{n_{l,r}\omega d}{c}$ is a phase shift.

We may express relations (\ref{pohybove rovnice}) in the normal coordinates
which are
%----------------------------------------------------------------------------
\begin{equation} \label{normalove souradnice}
\begin{array}{c}
q_{1} = \frac{1}{\sqrt{2(1+A_{1}^2)}}(r_{1}+A_{1}r_{2}+A_{1}r_{3}+r_{4}), \\
q_{2} = \frac{1}{\sqrt{2(1+A_{2}^2)}}(r_{1}+A_{2}r_{2}+A_{2}r_{3}+r_{4}), \\
q_{3} = \frac{1}{\sqrt{2(1+A_{3}^2)}}(r_{1}+A_{3}r_{2}-A_{3}r_{3}-r_{4}), \\
q_{4} = \frac{1}{\sqrt{2(1+A_{4}^2)}}(r_{1}+A_{4}r_{2}-A_{4}r_{3}-r_{4}),
\end{array}
\end{equation}
%----------------------------------------------------------------------------
where
%----------------------------------------------------------------------------
\begin{equation} \label{koeficienty v nor. sour.}
\begin{array}{c}
A_{1} =
\frac{Q_{1}+\sqrt{5Q_{1}^{2}+8Q_{1}Q_{2}+4Q_{2}^{2}}}{2(Q_{1}+Q_{2})}, \\
A_{2} =
\frac{Q_{1}-\sqrt{5Q_{1}^{2}+8Q_{1}Q_{2}+4Q_{2}^{2}}}{2(Q_{1}+Q_{2})}, \\
A_{3} =
\frac{Q_{1}-\sqrt{5Q_{1}^{2}-8Q_{1}Q_{2}+4Q_{2}^{2}}}{2(-Q_{1}+Q_{2})}, \\
A_{4} =
\frac{Q_{1}+\sqrt{5Q_{1}^{2}-8Q_{1}Q_{2}+4Q_{2}^{2}}}{2(-Q_{1}+Q_{2})}.
\end{array}
\end{equation}
%-----------------------------------------------------------------------------
As a result of coupling the natural frequency $\omega_{0}$ of each
oscillator splits into four characteristic frequencies of the normal modes of
vibrations
%----------------------------------------------------------------------------
\begin{equation} \label{normalove mody}
\begin{array}{c}
\omega_{1}^2 = \omega_{0}^2 + \frac{Q_{1}+\sqrt{5Q_{1}^{2}+8Q_{1}Q_{2}+
4Q_{2}^{2}}}{2}, \\
\omega_{2}^2 = \omega_{0}^2 + \frac{Q_{1}-\sqrt{5Q_{1}^{2}+8Q_{1}Q_{2}+
4Q_{2}^{2}}}{2}, \\
\omega_{3}^2 = \omega_{0}^2 - \frac{Q_{1}-\sqrt{5Q_{1}^{2}-8Q_{1}Q_{2}+
4Q_{2}^{2}}}{2}, \\
\omega_{4}^2 = \omega_{0}^2 - \frac{Q_{1}+\sqrt{5Q_{1}^{2}-8Q_{1}Q_{2}+
4Q_{2}^{2}}}{2}.
\end{array}
\end{equation}
%-----------------------------------------------------------------------------

Now it is possible to rewrite the motion equations of each oscillator
%-----------------------------------------------------------------------------
\begin{equation} \label{poh. rov. v norm. sour.}
\ddot q_{\eta}^{l,r}+\omega_{\eta}^{2}q_{\eta}^{l,r} =
R_{q_{\eta}}^{l,r},\qquad \eta =1,2,3,4,
\end{equation}
%-----------------------------------------------------------------------------
where $R_{q_{\eta}}^{l,r}$ are the forces in the normal coordinates. For
them we get in general expression
%-----------------------------------------------------------------------------
\begin{equation} \label{normalove sily oscilatoru}
R_{q_{\eta}}^{l,r} = \frac{eE_{0}}{m}(a_{q_{\eta}}^{l,r})
\cos(\omega t+\sigma _{q_{\eta }}^{l,r}), \qquad\eta = 1,2,3,4
\end{equation}
%------------------------------------------------------------------------
where $\sigma _{q_{\eta }}^{l,r}$ is a phase shift and for
$(a_{q_{\eta}}^{l,r})$ we have derived
%-------------------------------------------------------------------------
\begin{equation} \label{koef. v norm. silach}
\begin{array}{c}
(a_{q_{1}}^{l,r})^2 = \frac{\alpha^2+\beta^2}{1+A_{1}^{2}}[1 \pm
3\phi_{l,r} \sin\theta - 3\cos\theta + 4\cos^3\theta \mp
12\phi_{l,r}\cos^2\theta sin\theta  \\
- 2A_{1} (1 - \cos\theta -
2\cos^2\theta \pm \phi_{l,r}\sin\theta \pm 4\phi_{l,r}\sin\theta
\cos\theta)  \\
+ A_{1}^{2} (1 + \cos\theta \mp \phi_{l,r}\sin\theta)] \\
(a_{q_{2}}^{l,r})^2 =  \frac{\alpha^2+\beta^2}{1+A_{2}^{2}}[1 \pm
3\phi_{l,r} \sin\theta - 3\cos\theta + 4\cos^3\theta \mp
12\phi_{l,r}\cos^2\theta sin\theta  \\
- 2A_{2} (1 - \cos\theta -
2\cos^2\theta \pm \phi_{l,r}\sin\theta \pm 4\phi_{l,r}\sin\theta
\cos\theta)  \\
+ A_{2}^{2} (1 + \cos\theta \mp \phi_{l,r}\sin\theta)] \\
(a_{q_{3}}^{l,r})^2 = \frac{\alpha^2+\beta^2}{1+A_{3}^{2}}[1 \mp
3\phi_{l,r} \sin\theta + 3\cos\theta - 4\cos^3\theta \pm
12\phi_{l,r}\cos^2\theta sin\theta  \\
+ 2A_{3} (1 + \cos\theta -
2\cos^2\theta \mp \phi_{l,r}\sin\theta \pm 4\phi_{l,r}\sin\theta
\cos\theta)  \\
+ A_{3}^{2} (1 - \cos\theta \pm \phi_{l,r}\sin\theta)] \\
(a_{q_{4}}^{l,r})^2 = \frac{\alpha^2+\beta^2}{1+A_{4}^{2}}[1 \mp
3\phi_{l,r} \sin\theta + 3\cos\theta - 4\cos^3\theta \pm
12\phi_{l,r}\cos^2\theta sin\theta  \\
+ 2A_{4} (1 + \cos\theta -
2\cos^2\theta \mp \phi_{l,r}\sin\theta \pm 4\phi_{l,r}\sin\theta
\cos\theta)  \\
+ A_{4}^{2} (1 - \cos\theta \pm \phi_{l,r}\sin\theta)].
\end{array}
\end{equation}
%------------------------------------------------------------------------

Let's substitute eq. (\ref{normalove sily oscilatoru}) into eq.
(\ref{poh. rov. v norm. sour.}), we get
%-----------------------------------------------------------------------------
\begin{equation} \label{poh. rov. v norm. sour. 2}
\ddot q_{\eta}^{l,r}+\omega_{\eta}^{2}q_{\eta}^{l,r} = (a_{q_{\eta}}^{l,r})
\frac{eE_{0}}{m}\cos(\omega t+\sigma _{q_{\eta }}^{l,r})
\end{equation}
%------------------------------------------------------------------------------
and the solution of these equations is
%------------------------------------------------------------------------------
\begin{equation} \label{reseni pohybovych rovnic}
q_{\eta} = (a_{q_{\eta}}^{l,r})\frac{eE_{0}}{m}
\cdot \frac{\cos(\omega t+\sigma _{q_{\eta }}^{l,r})}{\omega_{\eta}^2-\omega^{2}}.
\end{equation}
%-----------------------------------------------------------------------------

The propagating light wave induces the dipole moments
$d_{q_{\eta}}^{l,r}$ which are
%-----------------------------------------------------------------------------
\begin{equation} \label{dipolove momenty}
d_{q_{\eta}}^{l,r} = q_{\eta}(a_{q_{\eta}}^{l,r})f_{q_{\eta}}e
\end{equation}
%-----------------------------------------------------------------------------
or using (\ref{reseni pohybovych rovnic})
%-----------------------------------------------------------------------------
\begin{equation} \label{dipolove momenty 2}
d_{q_{\eta}}^{l,r} = (a_{q_{\eta}}^{l,r})^2\frac{f_{q_{\eta}}e^2E_{0}}{m}
\cdot \frac{\cos(\omega t+\sigma _{q_{\eta }}^{l,r})}{\omega_{\eta}^2-\omega^{2}},
\end{equation}
%-----------------------------------------------------------------------------
where $f_{q_{\eta }}$ are the oscillator strengths in the normal modes
of vibrations. The mean polarizability per volume unit is
%-----------------------------------------------------------------------------
\begin{equation} \label{polarizovatelnost}
\chi_{q_{\eta}}^{l,r} = \frac{N'd_{q_{\eta}}^{l,r}}{E_{0}\cos(\omega
t +\sigma _{q_{\eta }}^{l,r})} = \frac{N
d_{q_{\eta}}^{l,r}}{3E_{0}\cos(\omega
t +\sigma _{q_{\eta }}^{l,r})},
\end{equation}
%----------------------------------------------------------------------------
where $N'$ is the number of compound oscilators in the volume unit. We see
that in our case $N'=N/3$ where $N$ is the number of isolated
oscillators.

For the refractive indices of the crystals we have the
Drude-Sellmaier dispersion relation
%----------------------------------------------------------------------------
\begin{equation} \label{disperzni vztah 2}
n_{l,r}^2-1 = 4\pi \sum_{\eta=1}^{4} \chi_{q_{\eta}}^{l,r} = \frac{4\pi
Ne^2}{3m} \sum_{\eta=1}^{4}
(a_{q_{\eta}}^{l,r})^2 \frac{f_{q_{\eta}}}{\omega_{\eta}^2-\omega^2}.
\end{equation}
%----------------------------------------------------------------------------
From eq. (\ref{disperzni vztah 2}) we are able to solve the
relation
%----------------------------------------------------------------------------
\begin{equation} \label{rozdil indexu}
n_{l}^2-n_{r}^2 = \frac{4\pi Ne^2}{3m} \sum_{\eta=1}^{4}
\frac{[(a_{q_{\eta}}^{l})^2-(a_{q_{\eta}}^{r})^2]
f_{q_{\eta}}}{\omega_{\eta}^2-\omega^2}
\end{equation}
%----------------------------------------------------------------------------
and using (\ref{koef. v norm. silach}) we can calculate for the expression
in square brackets
%-----------------------------------------------------------------------------
\begin{equation} \label{rozdily koeficientu}
\begin{array}{c}
(a_{q_{1}}^{l})^2-(a_{q_{1}}^{r})^2 =
\frac{(\alpha^2+\beta^2)(\phi_{l}+\phi_{r})}{1+A_{1}^{2}}\sin\theta
(3-12\cos^2\theta \\
-2A_{1}-8A_{1}\cos\theta-A_{1}^{2}), \\
(a_{q_{2}}^{l})^2-(a_{q_{2}}^{r})^2 =
\frac{(\alpha^2+\beta^2)(\phi_{l}+\phi_{r})}{1+A_{2}^{2}}\sin\theta
(3-12\cos^2\theta \\
-2A_{2}-8A_{2}\cos\theta-A_{2}^{2}), \\
(a_{q_{3}}^{l})^2-(a_{q_{3}}^{r})^2 =
\frac{(\alpha^2+\beta^2)(\phi_{l}+\phi_{r})}{1+A_{3}^{2}}\sin\theta
(-3+12\cos^2\theta \\
-2A_{3}+8A_{3}\cos\theta+A_{3}^{2}), \\
(a_{q_{4}}^{l})^2-(a_{q_{4}}^{r})^2 =
\frac{(\alpha^2+\beta^2)(\phi_{l}+\phi_{r})}{1+A_{4}^{2}}\sin\theta
(-3+12\cos^2\theta \\
-2A_{4}+8A_{4}\cos\theta+A_{4}^{2}).
\end{array}
\end{equation}
%-------------------------------------------------------------------------------

In eq. (\ref{rozdily koeficientu}) we can write $(\phi_{l}+\phi_{r}) =
\frac{\omega d(n_{l}+n_{r})}{c}$ and $n_{l}^2-n_{r}^2 =
(n_{l}-n_{r})(n_{l}+n_{r})$. Using the well known formula for the ORD
$\rho (\omega )=\frac{\omega }{2c}(n_{l}-n_{r})$ we
obtain
%-----------------------------------------------------------------------------
\begin{eqnarray} \label{rotacni polarizace 2}
%\begin{array}{c}
\rho(\omega) &=& \frac{2\pi
Nde^{2}}{3c^{2}}\omega^2(\alpha^{2}+\beta^{2})\sin\theta \nonumber\\
&&\times\biggl[\frac{(3-12\cos^2\theta-2A_{1}-8A_{1}\cos\theta-A_{1}^{2})f_{q_{1}}}
{(1+A_{1}^{2})(\omega_{1}^2-\omega^{2})}\nonumber\\
&&+\frac{(3-12\cos^2\theta-2A_{2}-8A_{2}\cos\theta-A_{2}^{2})f_{q_{2}}}
{(1+A_{2}^{2})(\omega_{2}^2-\omega^{2})} \nonumber\\
&&+\frac{(-3+12\cos^2\theta-2A_{3}+8A_{3}\cos\theta+A_{3}^{2})f_{q_{3}}}
{(1+A_{3}^{2})(\omega_{3}^2-\omega^{2})}\nonumber\\
&&+\frac{(-3+12\cos^2\theta-2A_{4}+8A_{3}\cos\theta+A_{4}^{2})f_{q_{4}}}
{(1+A_{4}^{2})(\omega_{4}^2-\omega^{2})}\biggr].
%\end{array}
\end{eqnarray}

Using eqs. (\ref{normalove mody}) we can rewrite the formula
(\ref{rotacni polarizace 2}) to the form which contains the natural
frequency of the oscillators only but we see that we would get the
complicated expression. We will discuss the result (\ref{rotacni
polarizace 2}) only in the approximations of the oscillator strengths
and also only for the crystals with the space groups of symmmetry
$D_{3}^{4}$ and $D_{3}^{6}$ for which $\theta =120$ deg and therefore
$\cos \theta =-1/2$.

%-----------------------------------------------------------------------------
\subsection{Discussion}
\label{disc}
\hspace \parindent
%-----------------------------------------------------------------------------
We will express as in the sections \ref{uvod}, \ref{kap2} and \ref{sec3} the
result (\ref{rotacni polarizace 2}) in the linear harmonic oscillator
approximation
and in the Heitler - London approximation of the oscillator strengths.
In all results we neglect all the small terms containing the expressions
$1/(\omega _{0}^{2}-\omega ^{2})$ of
the order higher than second. These terms contain with the
exception of the great value
of the denominator also the higher orders of the
small coupling constants $Q_{1}$ and $Q_{2}$ in the numerators.

At first we will use in the discussion of the
result (\ref{rotacni polarizace 2}) the Chandrasekhar linear harmonic
oscillator approximation where we can calculate with equation
$f_{{q}_{1}} =f_{{q}_{2}} =f_{{q}_{3}} =f_{{q}_{4}} = f_{0}$. In this case
we obtain the ORD formula of the Chandrasekhar type
%----------------------------------------------------------------------------
\begin{equation} \label{Chandrasekhar LHOA}
\rho(\omega) = \frac{4\pi Nde^2f_{0}(\alpha^2+\beta^2) \sin\theta
\cdot \left(Q_{1}-\frac{4}{3}Q_{2}\right) }{mc^2}
\cdot \frac{\omega ^{2}}{(\omega_{0}^{2}-\omega^{2})^{2}}.
\end{equation}
%----------------------------------------------------------------------------

The Heitler-London aproximation is in the case of four coupled
oscillators given by the relation
%-----------------------------------------------------------------------------
\begin{equation} \label{HLA}
\frac{f_{q_{1}}}{\omega_{1}} = \frac{f_{q_{2}}}{\omega_{2}} =
\frac{f_{q_{3}}}{\omega_{3}} = \frac{f_{q_{4}}}{\omega_{4}} =
\frac{f_{0}}{\omega_{0}}
\end{equation}
%-----------------------------------------------------------------------------
and using eqs. (\ref{normalove
mody}) and taking again into account that $Q_{1}$ and $Q_{2}$ are small
quantities we can simplify expressions for the oscillator strengths to
the form
%-----------------------------------------------------------------------------
\begin{equation} \label{sily oscilatoru HLA}
\begin{array}{c}
f_{q_{1}} = f_{0}(1+\frac{Q_{1}+\sqrt{5Q_{1}^{2}+8Q_{1}Q_{2}+
4Q_{2}^{2}}}{4\omega_{0}^{2}}), \\
f_{q_{2}} = f_{0}(1+\frac{Q_{1}-\sqrt{5Q_{1}^{2}+8Q_{1}Q_{2}+
4Q_{2}^{2}}}{4\omega_{0}^{2}}), \\
f_{q_{3}} = f_{0}(1-\frac{Q_{1}-\sqrt{5Q_{1}^{2}-8Q_{1}Q_{2}+
4Q_{2}^{2}}}{4\omega_{0}^{2}}), \\
f_{q_{4}} = f_{0}(1-\frac{Q_{1}+\sqrt{5Q_{1}^{2}-8Q_{1}Q_{2}+
4Q_{2}^{2}}}{4\omega_{0}^{2}}).
\end{array}
\end{equation}
%-----------------------------------------------------------------------------

Now we substitute these relations into eq. (\ref{rotacni polarizace 2})
and we obtain
%-----------------------------------------------------------------------------
\begin{eqnarray} \label{rotacni polarizace HLA}
%\begin{array}{c}
\rho(\omega) &=& \frac{2\pi
Nde^{2}f_{0}}{3mc^{2}}\frac{\omega^2}{\omega_{0}^{2}}(\alpha^{2}+\beta^{2})
\sin\theta \nonumber\\
&&\times\left[-\frac{3Q_{1}-4Q_{2}}
{\omega_{0}^2-\omega^{2}}+
\frac{2(3Q_{1}\omega_{0}^{2}-4Q_{2}\omega_{0}^{2}-2Q_{1}^{2}+2Q_{1}Q_{2})}
{(\omega_{0}^2-\omega^{2})^{2}} \right]
%\end{array}
\end{eqnarray}
%-----------------------------------------------------------------------------
and after neglecting the
terms $Q_{1}^{2}$ and $Q_{1}Q_{2}$ and adding the terms in the square
brackets we have the formula of the Agranovich type
%-----------------------------------------------------------------------------
\begin{equation} \label{Agranovich HLA}
%\begin{array}{c}
\rho(\omega) = \frac{2\pi
Nde^{2}f_{0}(\alpha^{2}+\beta^{2})\sin\theta \cdot (Q_{1}-\frac{4}{3}Q_{2})}
{mc^{2}}\cdot
\frac{\omega^{2}(\omega _{0}^{2}+\omega^{2})}{\omega
_{0}^{2}(\omega_{0}^2-\omega^{2})^2}.
%\end{array}
\end{equation}
%------------------------------------------------------------------------------

We see that applying the linear harmonic oscillator aproximation and
the Heitler-London
aproximation to the general ORD formula (\ref{rotacni polarizace 2})
of four coupled oscillator model we obtained for the crystals belonging to
the space groups of symmetry $D_{3}^{4}$ and $D_{3}^{6}$ the Chandrasekhar
(\ref{Chandrasekhar LHOA}) and the Agranovich (\ref{Agranovich HLA})
formulae again. Both formulae contain the same expression
$Q_{1}-\frac{4}{3}Q_{2}$. Let's also note the results of solving
three coupled oscillator model published in \cite{Vys5} where these formulae
contain the expression $Q_{1}-Q_{2}$.

We obtain the similar results also in the five, six etc. oscillator
models. In the linear harmonic oscillator and in the Heitler - London
approximation we get again the Chandrasekhar and the Agranovich formulae
which are different only in the constant containing the coupling
constants $Q_{1}$ and $Q_{2}$. This constant has for the five oscillator
model the form $Q_{1}-\frac{3}{2}Q_{2}$, in the six oscillator model the
form $Q_{1}-\frac{8}{5}Q_{2}$ etc.

These results give the possibility to suppose the form of the ORD
formulae for general $\bar N$ coupled oscillator model. We obtain the
Chandrasekhar formula in the form
%-----------------------------------------------------------------------------
\begin{equation} \label{Chandrasekhar N}
\rho(\omega) = \frac{4\pi Nde^2f_{0}(\alpha^2+\beta^2) \sin\theta
\cdot \left( Q_{1}-\frac{2({\bar N}-2)}{{\bar N}-1}Q_{2}\right) }{mc^2}
\cdot \frac{\omega ^{2}}{(\omega_{0}^{2}-\omega^{2})^{2}},
\end{equation}
%-----------------------------------------------------------------------------
where ${\bar N}-1$ is the number of coupling between neighbouring oscillators
and ${\bar N}-2$ the number of coupling between odd and even oscillators
in the model.
It means that for the practical case ${\bar N}\rightarrow\infty$ we have
%-----------------------------------------------------------------------------
\begin{equation} \label{Chandrasekhar nekonecno}
\rho(\omega) = \frac{4\pi Nde^2(\alpha^2+\beta^2) \sin\theta
\cdot \left(Q_{1}-2Q_{2}\right) }{mc^2}
\cdot \frac{\omega ^{2}}{(\omega_{0}^{2}-\omega^{2})^{2}}.
\end{equation}
%-----------------------------------------------------------------------------
The same results we can write for the Agranovich formula:
%-----------------------------------------------------------------------------
\begin{equation} \label{Agranovich N}
\rho(\omega) = \frac{2\pi
Nde^{2}f_{0}(\alpha^{2}+\beta^{2})\sin\theta \cdot (Q_{1}-\frac{2({\bar
N}-2)}{{\bar N}-1}Q_{2})}
{mc^{2}}
\cdot \frac{\omega^{2}(\omega_{0}^{2}+\omega^{2})}{\omega
_{0}^{2}(\omega_{0}^2-\omega^{2})^2}
\end{equation}
%-----------------------------------------------------------------------------
and for ${\bar N}\rightarrow \infty $
\begin{equation} \label{Agranovich N1}
\rho(\omega) = \frac{2\pi
Nde^{2}f_{0}(\alpha^{2}+\beta^{2})\sin\theta \cdot(Q_{1}-2Q_{2})}
{mc^{2}}
\cdot \frac{\omega^{2}(\omega_{0}^{2}+\omega^{2})}{\omega
_{0}^{2}(\omega_{0}^2-\omega^{2})^2}.
\end{equation}
%-----------------------------------------------------------------------------

We see that the expression $Q_{1}-2Q_{2}$ is the common limit of the
coefficients containing the coupling constants in the final ORD formulae
in the case ${\bar
N}\rightarrow \infty $.
%-----------------------------------------------------------------------------

\section{Conclusions}
\label{conc}
\hspace \parindent
In this paper we have proved that with the Chandrasekhar ORD two coupled
oscillator model we can include all couplings between adjacent
oscillators on the helix that is given by the crystal structure. We have
solved that the influence of the couplings between adjacent oscillators
in the cases when the second single oscillator in the first
compound oscillator is the first oscillator in the second compound
oscillator
we can hold as aditive (the mistake in this conclusion is
neglectable). For the including of all couplings between adjacent
oscillators is sufficient to take in the Chandrasekhar model that the
number of single oscillators in the volume unit is the same as the
number of the couplings between adjacent oscillators. The same results
were proved for the models of three and four coupled oscillators - the
parts of the ORD results that contain the coupling constant between
adjacent oscillators $Q_{1}$ are identical in all models.

From the crystal structure it follows that we cannot neglect the
couplings between even and odd oscillators on the helix.
We can assume for the real crystal that the value of the coupling
constant between even
and odd oscillators $Q_{2}$ is comparable with the value of the coupling
constant between adjacent oscillators $Q_{1}$. Further we have proved
that in the crystals with the space groups of symmetry $D_{3}^{4}$ and
$D_{3}^{6}$ the couplings between even and odd oscillators have the
opposite effect on the ORD than the couplings between adjacent
oscillators.

The first coupled oscillator model in which we can include the coupling
between even or odd oscillators is the three oscillator model. But
using the previous conclusion this model can include all couplings
between adjacent oscillators and only a half of couplings between even
and odd oscillators on the helix. Besides we cannot assume these
couplings as aditive because they function between single oscillators
inside of the compound oscillators - for example in the three oscillator
model the second oscillator in the compound oscillator is in the
fact coupled also with the second oscillator in the adjacent compound
oscillator (becouse in the used model the last single oscillator in
any compound oscillator is the first single oscillator in the adjacent
compound oscillator).

From this reason we have further solved the models of four, five etc.
coupled oscillators. In these models the number of the included couplings
between even and odd oscillators increases and the results can
be generalized. By this way we have proved that the couplings
between even and odd oscillators also have the aditive effects. The ORD
results of the three oscillators model after the approximations of the
oscillator strengths of the normal modes of vibrations contain the
coefficient $Q_{1}-Q_{2}$ (for the crystals with space groups of
symmetry $D_{3}^{4}$ and $D_{3}^{6}$). In the limit of $\bar N$ coupled
oscillators for ${\bar N}\rightarrow \infty $ it was derived the
coefficient $Q_{1}-2Q_{2}$. This also means that the relative influence
of the couplings between even and odd oscillators on the ORD is twofold
in comparison with the couplings between adjacent oscillators.

We can formally obtain the generalized ORD results of the $\bar N$ coupled
oscillators model after the approximations of the oscillator strengths
using the three oscillators model in which we denote the
value of coupling between the first and the third oscillators as
$2Q_{2}$. We can say that this conclusion is practically acceptable
only after using the
results of the three oscillator model in the numerical approximations
of the experimental ORD data for example of $\alpha $-quartz or tellur.

\label{konec-vys4}
\end{document}